\documentclass[apj]{emulateapj}
\usepackage{graphicx}
\begin{document}
\submitted{Accepted for publication in the Astrophysical Journal}

\title{Very cold and massive cores near ISOSS J18364-0221:\\
Implications for the initial conditions of high-mass star-formation}

\shorttitle{Very cold and massive cores near ISOSS J18364-0221}
\shortauthors{Birkmann, Krause \& Lemke}

\author{Stephan M. Birkmann\altaffilmark{1}}\email{birkmann@mpia.de}
\author{Oliver Krause\altaffilmark{1,2}}\email{krause@as.arizona.edu}
\author{Dietrich Lemke\altaffilmark{1}}\email{lemke@mpia.de}

\altaffiltext{1}{Max-Planck-Institut f\"ur Astronomie, K\"onigstuhl 17, D-69117 Heidelberg, Germany}
\altaffiltext{2}{Steward Observatory, University of Arizona, Tucson, AZ 85721}

\begin{abstract}
  We report the discovery of two very cold and massive molecular cloud cores
  in the region ISOSS J18364-0221. The object has been identified by a
  systematic search for very early evolutionary stages of high-mass stars using
  the 170\,$\mu$m ISOPHOT Serendipity Survey (ISOSS). Submm continuum and molecular line
  measurements reveal two compact cores within this region. The first core has a 
  temperature of 16.5\,K, shows signs of ongoing infall and outflows, has no NIR or MIR
  counterpart and is massive enough ($M \sim 75$\,M$_\odot$) to form at least
  one O star with an associated cluster. It is therefore considered a candidate for
  a genuine high-mass protostar and a high-mass analog to the Class 0 objects. The
  second core has an average gas and dust temperature of only $\sim 12$\,K and a mass
  of $M \sim 280$\,M$_\odot$. Its temperature and level of turbulence are
  below the values found for massive cores so far and are suggested
  to represent the initial conditions from which high-mass star formation occurs.
\end{abstract}


%

\section{Introduction}
Most of the current knowledge about the formation and early evolution 
of high-mass stars has been inferred from observations of luminous 
infrared sources: High-mass protostellar objects (HMPO) \citep{sridharan}
and hot molecular cores (HMCs) \citep{kurtz}, both among the youngest known 
manifestations of high-mass stars, are dense condensations of warm (T $> 30$\,K)
gas and dust.
Given their high luminosity of $L_\mathrm{IR} > 10^3$\,L$_\odot$ most of 
these objects were readily detected by the IRAS mission and have been 
identified by using color criteria on their infrared fluxes
\citep[e.g.][]{fontani,williams,molinari,henning}. 

Much less is known about a potential evolutionary stage prior to the HPMO/HMC 
phase. It has been suggested that very cold (T $\sim$ 10\,K) and massive cores 
might represent the initial conditions from which massive stars form \citep{evans},
similar to the pre-stellar cores in the low-mass regime \citep{ward}. Although promising 
candidates for very young protostars have been recently discovered 
near more evolved and luminous objects \citep{garay,forbrich,wu}, 
no clear example of such a very cold core is known so far.

Using the 170\,$\mu$m ISOPHOT Serendipty Survey (ISOSS) \citep{lemke,bogun} we have
performed an unbiased search for such very cold sources based on their
$F_\mathrm{170\mu m}/F_\mathrm{100\mu m}$ color.
By cross-correlating the ISOSS catalog with the IRAS point source catalog and 
molecular line surveys, we have identified more
than 50 massive ($M > 100$\,M$_\odot$) candidate sources with a flux ratio of
$F_\mathrm{170\mu m}/F_\mathrm{100\mu m} > 2$, implying average dust temperatures
lower than 18\,K \citep{krause04}.

We have launched multi-wavelength follow-up observations to explore the physical 
conditions in this unique sample of cold and massive star forming regions 
\citep[e.g.][]{krause03a}. In this letter we report the discovery of extremely cold 
and massive cloud cores in the region ISOSS J18364-0221 which provide new
constraints on the physical conditions under which high-mass stars form.

\section{Observations}
\begin{figure}
  \centering
  \plotone{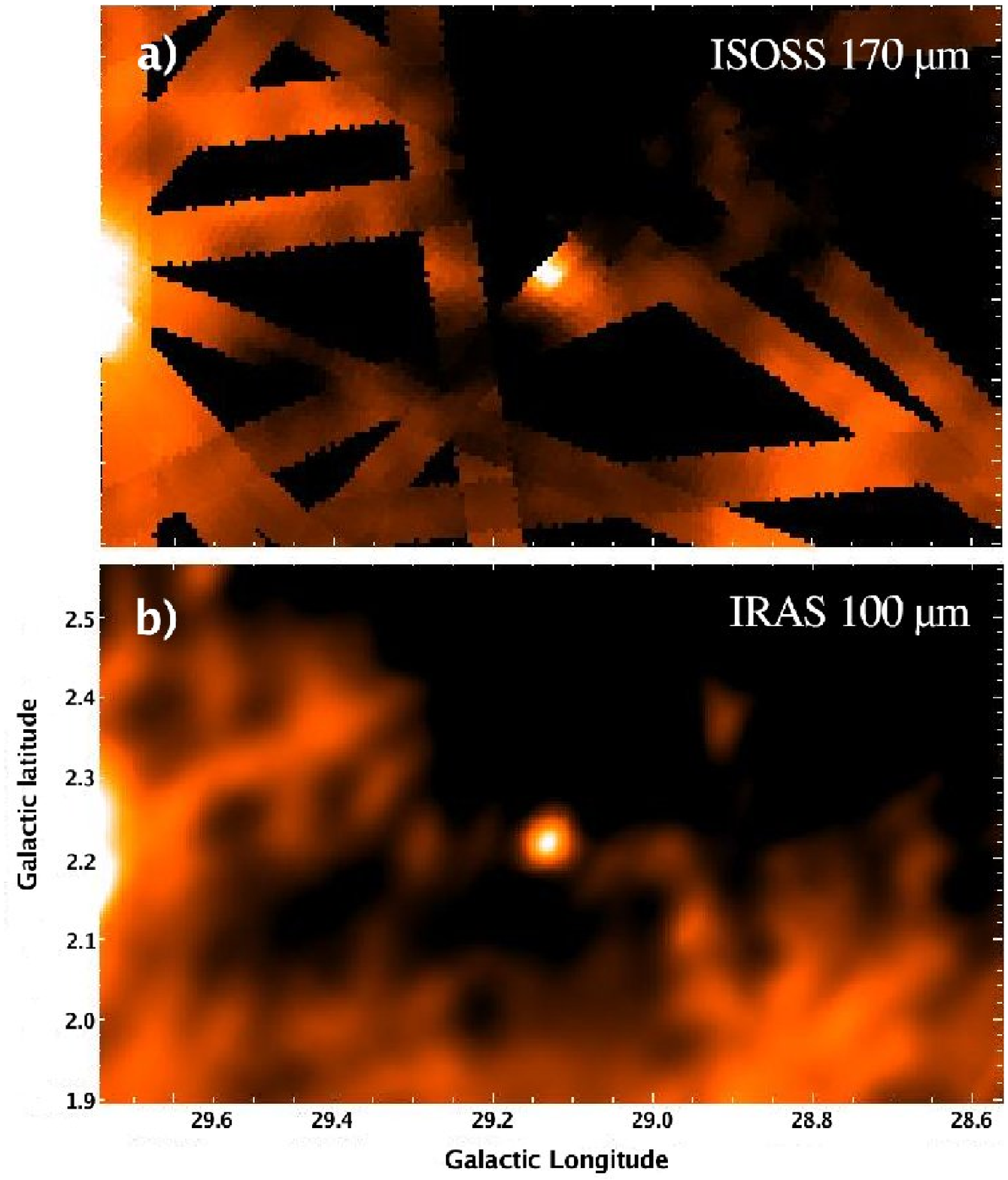}
  \caption{{\bf a)} Incomplete 170\,$\mu$m map constructed from slew-measurement by the
    ISOPHOT Serendipty Survey. ISOSS J18364-0221 is the compact source in the
    center. {\bf b)} 100\,$\mu$m map of the same field, produced by HIRES processing
    of the IRAS survey.
  }
  \label{fig:iso}
\end{figure}

Fig.~\ref{fig:iso} shows far-infrared observations of ISOSS J18364-0221.
The object was centrally crossed by two detector pixels of the C200 camera onboard 
ISO (beam size of 1.5\arcmin\,$FWHM$) during an ISOSS scan. A 170\,$\mu$m flux 
$F_\mathrm{170\mu m} = 235 \pm 47$\,Jy has been derived using procedures by
\citet{krause03} and \citet{stickel}. The compact ISOSS source coincides in 
position with the unresolved IRAS point source 18339-0224  
(R.A. (2000)=18$^\mathrm{h}$36$^\mathrm{m}$24\fs 7, DEC. (2000)=-02\degr 21\arcmin 49\arcsec) 
which has previously been detected in the CO(1-0) transition \citep{yang}. Its CO radial
velocity of $v_\mathrm{LSR}\sim 33$\,km/s corresponds to a kinematical distance
of $d \sim 2.2$\,kpc according to \citet{brand}. This is consistent with the distance
derived by the empirical relation $d = 320N^{0.57}$\,pc, where $N$ is the number of
foreground stars detected on the POSS-B image inside a 5\arcmin\ aperture towards the
cloud \citep{herbst}.

Submm continuum jiggle maps at 850\,$\mu$m and 450\,$\mu$m were obtained
with SCUBA \citep{holland} at the James Clerk Maxwell Telescope (JCMT) in
May 2003 under good atmospheric transmission conditions ($\tau_{225\mathrm{GHz}} =
0.067 \pm 0.003$).
The data was reduced with SURF, including sky noise reduction, yielding a photometric
accuracy of 20\% (450\,$\mu$m) and 12\% (850\,$\mu$m). The beam size is 7.5\arcsec{} and
14.5\arcsec, respectively.

Observations of the Ammonia inversion lines (J,K) = (1,1), (2,2), (3,3), and (4,4)
were carried out at the Effelsberg 100\,m telescope in January 2003. The beam
size at the observing frequency of 23.7\,GHz is 40\arcsec.
The pointing accuracy is estimated to 5\arcsec{} rms and the flux uncertainty to 10\%.

Molecular line measurements of CO(2-1) and C$^{18}$O(2-1) (beam size $\sim 11\arcsec$), 
HCO$^+$(3-2) and H$^{13}$CO$^+$(3-2) (beam size $\sim 9.5\arcsec$) were obtained at the 
IRAM 30\,m in November 2003 and of CO(3-2) (beam size $\sim 22\arcsec$) at the
Heinrich Hertz Telescope in December 2004. We used the facility receivers and
reduced all spectra with the CLASS software.
The atmospheric transmission wwas ($\tau_{225\mathrm{GHz}}\sim 0.15$) for the HCO$^+$(3-2)
($\tau_{225\mathrm{GHz}}\sim 0.5$) for the CO(2-1) and ($\tau_{225\mathrm{GHz}}\sim 0.1$)
for the CO(3-2) observations. The estimated pointing error for both telescopes is 3\arcsec{} 
and the radiometric accuracy 15\%.

Near infrared images in J, H, and Ks have been obtained at the
Calar Alto 3.5\,m telescope in Oktober 2003 and June 2004 with Omega2000
and OmegaPrime. Data reduction was carried out using standard procedures in IRAF.
Photometric calibration is based on the 2MASS point source catalog.


\section{Results}

Fig.~\ref{fig:submm} displays contours of SCUBA 850\,$\mu m$ emission
towards ISOSS J18364-0221 overlayed onto a near-infrared color composite.
The IRAS and ISOSS FIR point source is clearly resolved into two emission peaks in the
submillimeter. The $FWHM$ is $17\arcsec \times 10.5\arcsec$ $(\mathrm{PA} \sim -10\degr)$ 
for the eastern core (SMM1), while the western core (SMM2) is more extended.
The fluxes are $F_\mathrm{850\mu m} = 2.11$\,Jy and $F_\mathrm{450\mu m} = 13.5$\,Jy for
SMM1 and $F_\mathrm{850\mu m} = 2.85$\,Jy and $F_\mathrm{450\mu m} = 15.8$\,Jy for SMM2.
In addition to the photometric uncertainties of the submm fluxes, the 850\,$\mu$m
flux may be contaminated by the CO(3-2) transition. We checked the CO(3-2) contribution
in the SCUBA 850\,$\mu m$ band using our HHT CO(3-2) map and found the level to
be within our overall calibrational uncertainty.
Derivation of the dust temperature from submm data alone heavily depends on the
chosen dust emissivity index $\beta$ for which values between 1 and 2 are reported
in the literature. Using an emissivity index of $\beta = 2$ we find
$T_\mathrm{d} = 14$\,K for SMM1 and $T_\mathrm{d} = 12$\,K for SMM2. For lower
values of $\beta$ the temperatures would be higher. In order to constrain both
$\beta$ and the dust temperature we use the measured 100\,$\mu$m flux
($F_\mathrm{100\mu m} = 57.7$\,Jy) of the whole ISOSS J18364-0221
cloud that places an upper limit for the emission of the individual cores.
Using HIRES processed maps of the region, we find that the relative flux ratio
of SMM1 to SMM2 at 100\,$\mu$m is approximately 10:1. Using this ratio we derive
$\beta = 1.6$ and $T_\mathrm{d} = 21$\,K for SMM1 and $\beta = 1.8$ and
$T_\mathrm{d} = 13.8$\,K for SMM2. Since this is an upper limit we take the
average values for the rest of the paper: $\beta = 1.8\pm 0.2$, $T_\mathrm{d}
= 16.5^{+6}_{-3}$\,K and $\beta = 1.9\pm 0.1$, $T_\mathrm{d} = 12.8^{+2.0}_{-1.5}$\,K
for SMM1 and SMM2, respectively.

We estimate the masses of the cores using the relation of
\citet{hildebrand} $M_\mathrm{d} = F_\lambda d^2 /
\kappa_\lambda B_\lambda(T)$ with a gas-to-dust ratio of 100. The mass of SMM1 is
$M = 75\pm 30$\,M$_\odot$, SMM2 has a mass of $M = 280^{+75}_{-60}$\,M$_\odot$.
We used $\kappa_\mathrm{850\mu m} = 0.018$ for SMM1 (gas+dust, thin icemantles,
$n \sim 10^6$\,cm$^{-3}$) and for SMM2 $\kappa_\mathrm{850\mu m} = 0.010$ (gas+dust,
thin icemantles) taking the values from \citet{ossenkopf}).

\begin{figure}
  \centering
  \plotone{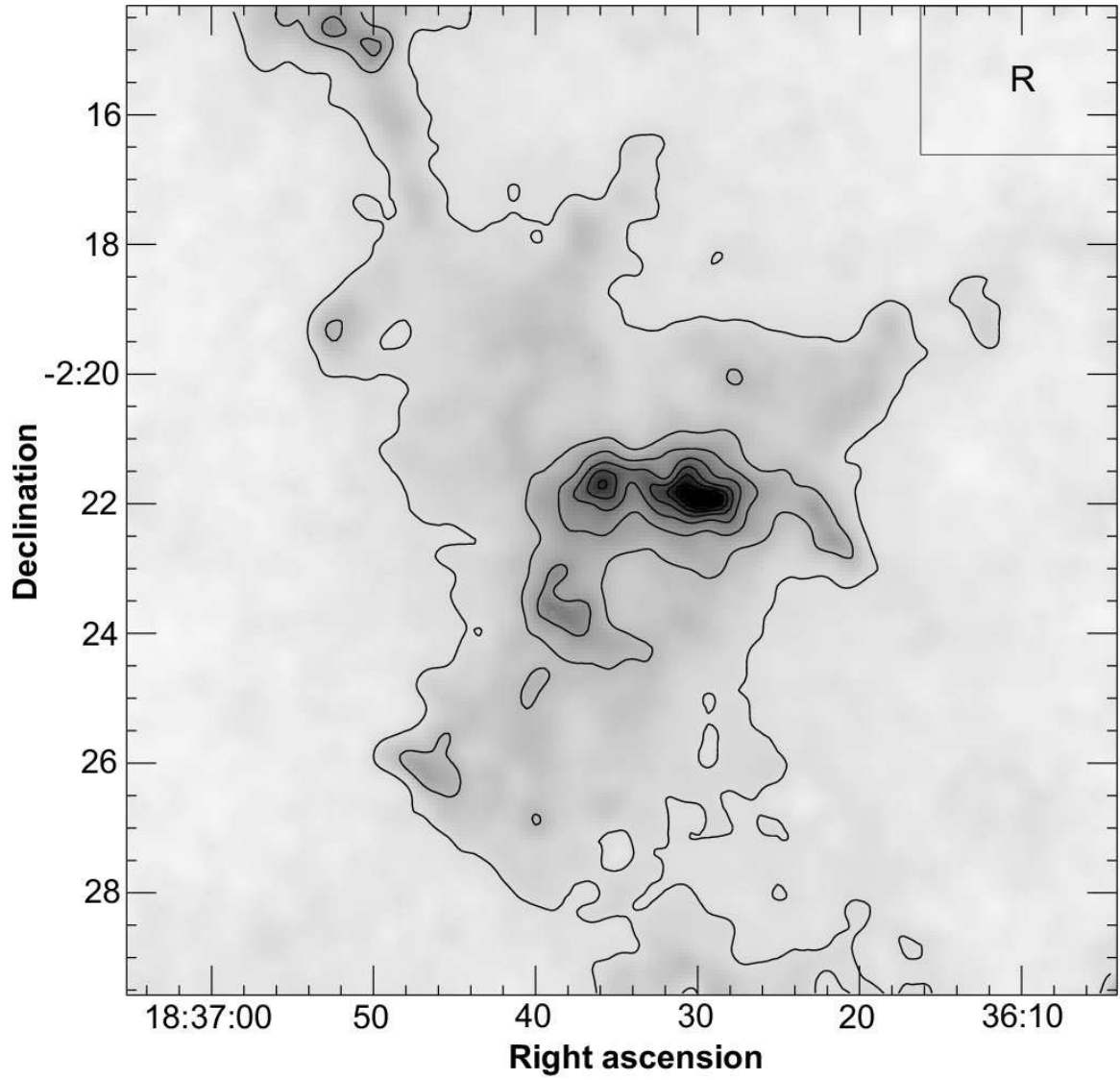}
  \caption{Map of the visual extinction $A_\mathrm{V}$ towards ISOSS J18364-0221.
    Contours start at $A_\mathrm{V}=4$ and increase in steps of 4. The local extinction was
    determined from the reference region R. Colors for zero extinction were established
    using a simulated besancon model catalog \citep{besancon} towards the region.
  }
  \label{fig:ext}
\end{figure}

The submillimeter emission is highly correlated with the NIR extinction, 
as seen by the reddening and surface density of background stars in Fig.~\ref{fig:submm}.
We have constructed a quantitative extinction map from our J, H, and Ks images according
to the \textsc{Nicer} method of \citet{lombardi} (Fig.~\ref{fig:ext}).
The map reveals that the compact far-infared and submm sources are
embedded in the center of a much more extended cloud complex which is
also visible as an optical dark cloud Lynds 541.

\begin{figure}
  \centering
  \plotone{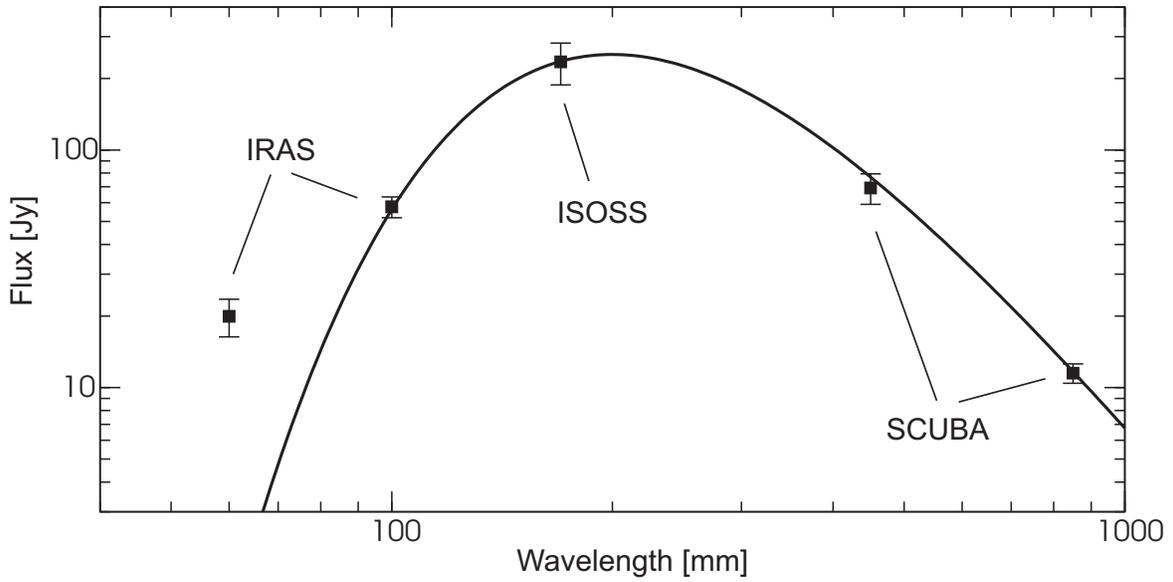}
  \caption{Spectral energy distribution of ISOSS J18364-0221. For wavelengths
    longwards of 100\,$\mu$m the SED is dominated by the optically thin thermal
    emission of large grains and well fitted with a modified black body with an
    emissivity index of $\beta = 2$, the point at 60\,$\mu$m is not taken into
    account.
  }
  \label{fig:sed}
\end{figure}

In contrast to the submillimeter maps which resolve individual features on 
typical scales of individual cloud cores (0.1\,pc), the ISOSS and IRAS 
beams constrain the average temperature and mass of the whole dense cloud center 
on approximately 1\,pc scale.
The small chopping throw of SCUBA (2\arcmin) results in spatial filtering of
structures that are comparable or larger in size. We account for an average
column density of $A_\mathrm{V} \sim 7$ (see Fig.~\ref{fig:ext}) to correct
the integrated flux of the chopped SCUBA measurements. We fit a single
modified black body to the data longwards of 100\,$\mu$m (Fig.~\ref{fig:sed})
and derive a temperature of $14.6^{+1.2}_{-1.0}$\,K and $\beta = 2$. The total
luminosity of ISOSS J18364-0221 is $L \sim 800$\,L$_\sun$.

The mass of the region is $M = 900^{+450}_{-330}$\,M$_\sun$
($\kappa_\mathrm{170\mu m} = 25$\,cm$^2$g$^{-1}$) with the uncertainty dominated by
the derived average dust temperature.
A lower limit for the mass can be computed from our extinction map, using the relation
$M = (\Delta \Omega d)^2 \mu \frac{N_\mathrm{H}}{A_\mathrm{V}} \sum_i
A_\mathrm{V}(i)$ given by \citet{dickman}. With a dust-to-gas ratio of
$N_\mathrm{H}/A_\mathrm{V} = 1.87\times 10^{21}$\,cm$^{-2}$\,mag$^{-1}$
\citep{savage} we obtain a mass of $M \geq 460$\,M$_\odot$
for ISOSS J18364-0221 and $M \sim 3200$\,M$_\odot$ for the whole complex
shown in Fig.~\ref{fig:ext}.

Ammonia (NH$_3$) is an ideal temperature probe for the molecular
gas in dense regions \citep{harju,walmsley}. From the relative
intensity of the (J,K) = (1,1) and (2,2) inversion transitions of our ammonia
(NH$_3$) data, we derive a gas kinetic temperature of $T_\mathrm{kin} = 11.6 \pm 1.5$\,K
towards SMM2 (Fig.~\ref{fig:nh3}). The inversion transitions (J,K) = (3,3) and (4,4)
were not detected in neither of the cores ($\sigma_{\rm RMS} \sim$ 30\,mK).
The latter transitions have been used as sensitive tracers of HMCs and argue against
the presence of such an object in ISOSS J18364-0221. SMM2 has a shallow
density profile and is larger than the beam size of the ammonia measurements (40\arcsec).
Therefore beam dillution is neglectable and we can directly compare the line and
continuum measurements. The good match of $T_\mathrm{kin}$ and $T_\mathrm{d}$
(in particular for $\beta = 2$) suggests thermal coupling between dust and gas,
as it is expected for cold and dense cores.


\section{Discussion}
\begin{figure}
  \centering
  \plotone{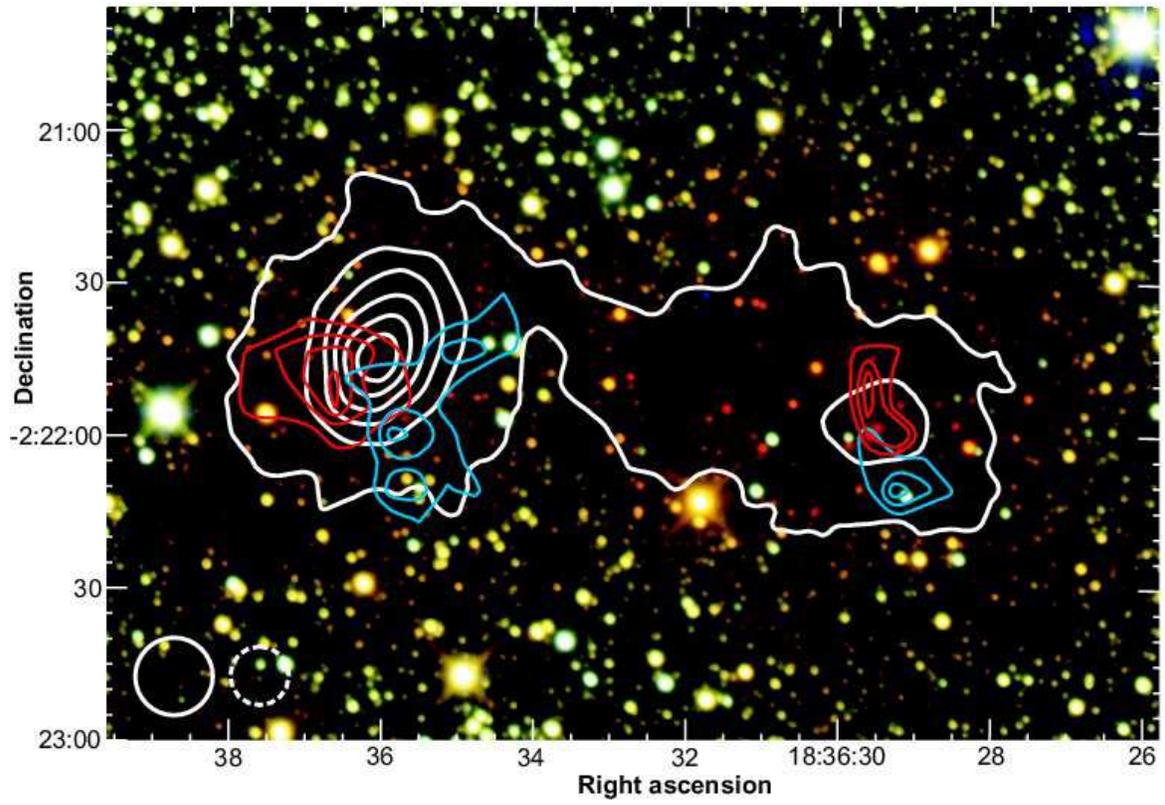}
  \caption{ISOSS J18364-0221 in JHKs, the white contours give the 850\,$\mu$m
    flux at levels of  50, 150, 250, $+200$\,mJy/beam. The thin contours
    show the red ($v[40-52]$\,km\,s$^{-1}$) and blue ($v[22-30]$\,km\,s$^{-1}$)
    CO(2-1) emission. The beam sizes are denoted in the lower left.
  }
  \label{fig:submm}
\end{figure}

\begin{figure}
  \centering
  \plotone{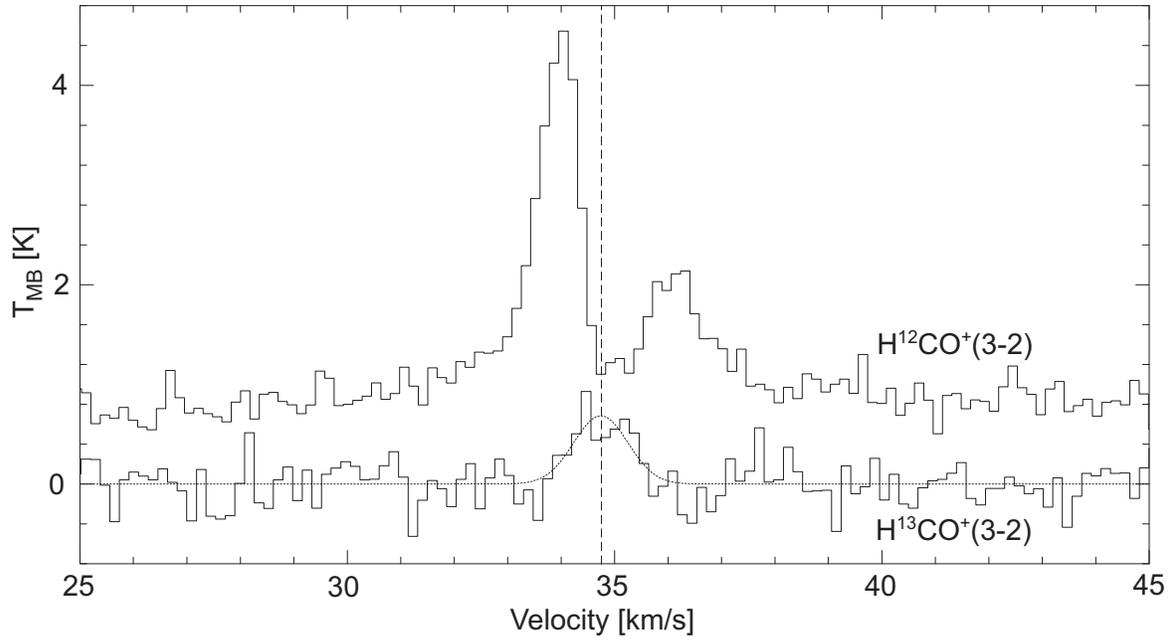}
  \caption{H$^{12}$CO$^+$(3-2) and H$^{13}$CO$^+$(3-2) taken at SMM1. The vertical
    dashed line marks the central velocity of the optical thin line, the
    optically thick transition shows red-shifted self absorption.
  }
  \label{fig:infall}
\end{figure}

We now discuss the nature of the two massive cold cores in
ISOSS J18364-0221 and the implications for the initial conditions
of high-mass star formation.\\
{\bf SMM1} is cold ($T \approx 16.5$\,K) and compact (effective
radius $R\sim 0.2$\,pc). From our 450\,$\mu$m data we derive a column density of
$N(\mathrm{H_2}) = 2.2\times 10^{23}$\,cm$^{-2}$ and a minimum central
density of $n(\mathrm{H_2}) = 1.4\times 10^6$\,cm$^{-3}$.
Our measurement of optically thin H$^{13}$CO$^+$(3-2) transition towards 
SMM1 (Fig.~\ref{fig:infall}) provides an independent check for the temperature and 
(column) density of the high-density gas. Using the density und temperature from 
the dust continuum and applying the radiative transfer code from \citet{schoeier},
we find an abundance of [HCO/H$_2$] of $8 \times 10^{-10}$ which is in good
agreement with the value predicted for protostellar cores \citep{bergin}.
Our molecular line measurements further indicate that the
core is collapsing and drives an outflow: The optically thick H$^{12}$CO$^+$(3-2) 
transition shows a red-shifted self absorption with
respect to the optically thin H$^{13}$CO$^+$(3-2) line (Fig.~\ref{fig:infall}), suggesting
large scale infall towards the center of the core \citep{choi,evans99}.
The CO(2-1) spectra show significant line wings around the two submm cores when compared with
C$^{18}$O(2-1), indicating the presence outflows (see Fig.~\ref{fig:submm}).
We derived the outflow parameters using our CO(2-1) and C$^{18}$O(2-1) data
following \citet{henning2}. The mass of the outflow is estimated to
$M_\mathrm{f} \approx 18$\,M$_\sun$ and the dynamical timescale $t_\mathrm{d}$ is
$\sim 1.8 \times 10^4$\,years. The corresponding mass outflow rate is
$\dot M \approx 1 \times 10^{-3}$\,M$_\sun$/yr and the momentum supply rate
$F \approx 8.5 \times 10^{-3}$\,M$_\sun\:\mathrm{km\:s^{-1}}$/yr,
without correcting for the unknown inclination $i$ of the outflow. This
correction would further increase the derived values, which are already comparable to
those observed for outflows in high-mass star forming regions
\citep[e.g.][]{beuther,zhang}. It must be noted however that
both the red and blue lobes have complex morpholgies and higher resolution observations
are required to decide on the actual number of outflows driven by SMM1.
No infrared counterpart is detected in our deep near-infrared images nor in the
MSX mid-infrared data, which is however not surprising considering the extinction of
$A_\mathrm{V} \geq 120$. With $M \sim 75$\,M$_\sun$ SMM1 would be massive enough to form an
O type star and an associated cluster. We therefore consider SMM1 as a promising candidate
for a massive protostar during its initial graviational collapse,
i.e. in an evolutionary state comparable to the Class 0 phase in low-mass star formation.

\begin{figure}
  \centering
  \plotone{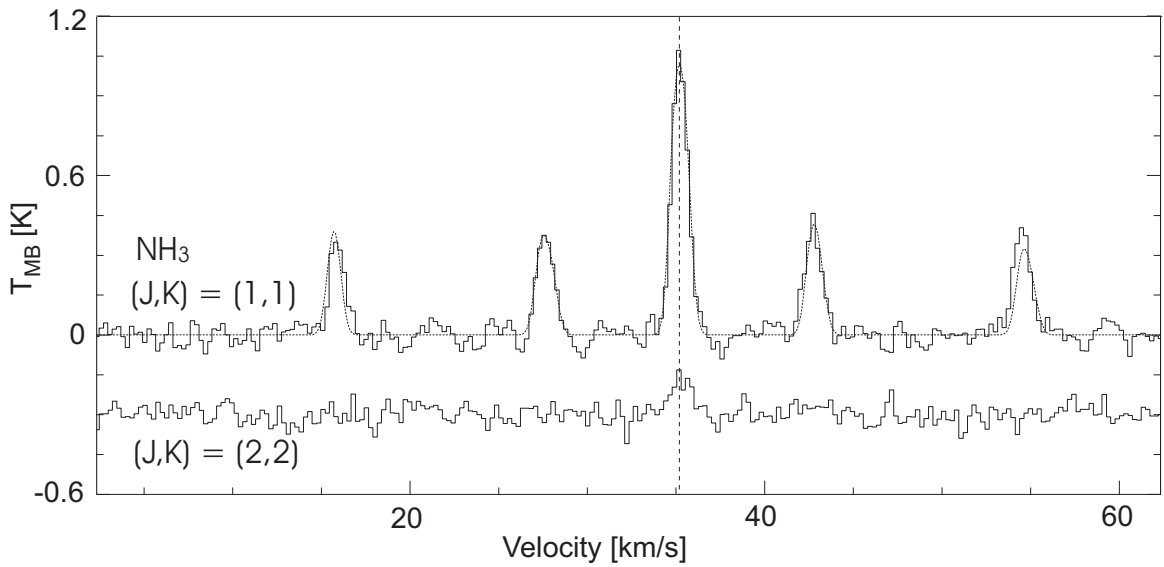}
  \caption{Spectra of the NH$_3$ (J,K) = (1,1) and (2,2) inversion transitions
    towards ISOSS J18339-0221 SMM2. A fit to the hyperfine structure of the
    (1,1) line is also shown.
  }
  \label{fig:nh3}
\end{figure}

{\bf SMM2} has very low gas and dust temperatures
($T \approx 12$\,K), moderate densities (central density
$n(\mathrm{H_2})\sim 2.3\times 10^{5}$cm${^-3}$), and is quite extended
(effective radius $R \sim 0.5$\,pc). Such conditions
have been suggested to characterize the earliest stage of star formation \citep{evans},
but have so far only been observed in low-mass prestellar cores.
We investigate the dynamical state of the core considering the virial theorem,
with $E_\mathrm{mag}+E_\mathrm{pot} = 2(E_\mathrm{kin}-E_\mathrm{ext})$ being the
condition for a gravitationally bound core. $E_\mathrm{mag}$ denotes the
magnetic energy, $E_\mathrm{pot}$ is the potential energy, $E_\mathrm{kin}$ the
total kinetic energy, and $E_\mathrm{ext}$ accounts for external pressure. The
kinetic energy can be written as $E_\mathrm{kin} = E_\mathrm{therm} +
E_\mathrm{turb} = \frac{3}{2} NkT + \frac{3}{2}M\sigma^2_{turb}$, with $\sigma^2
= \Delta V^2/(8 \ln 2) - kT/m$, where $\Delta V = 0.90$\,km/s is the linewidth taken
from our ammonia data (Fig.~\ref{fig:nh3}) and $m$ the mass of the NH$_3$ molecule. 
The potential energy of a homogenous ellipsoid is $E_\mathrm{pot} = 3GM^2/5R$ \citep{lilje}.
We find $E_\mathrm{pot} = 9.5\times 10^{38}$\,J, $E_\mathrm{therm} = 3.1\times 10^{37}$\,J,
and $E_\mathrm{turb} = 1.2\times 10^{38}$\,J. Neglecting the external pressure
and the magnetic field energy the total kinetic energy sums up to $\sim 16$\%
of the potential energy. We conclude that SMM2 is Jeans unstable.
The small line width of the optically thin ammonia transitions suggests that the level 
of turbulence is not as high as in more evolved high-mass cores. There are two sources
with very red colors towards SMM2 (H-K=3.7 and H-K=4.5) that are presumably low-mass
Class I objects, based on their NIR and MIR fluxes with a spectral index of
$a = \mathrm{d}\log(\lambda F_\lambda) / \mathrm{d} \log(\lambda) \sim 3$. This
classification is supported by the molecular outflow traced in CO(2-1) emerging
from the young stellar objects (Fig.~\ref{fig:submm}). This supports the 
large-scale Jeans instability of the core and indicates that fragmentation and 
cluster formation has already started. Therefore SMM2 is a very promising
candidate for a massive pre-stellar core, characterizing the earliest phase of star
formation.

Several authors \citep[e.g.][]{evans} have suggested that the
conditions during the earliest phases of high-mass star-formation should be similar 
to that prevailing in the regime of low-mass star birth. Our findings provide direct
observational evidence that the initial conditions of high-mass star-formation
are indeed characterized by very low temperatures and low levels of turbulence.


\begin{acknowledgements}
We like to thank the anonymous referee for the helpful comments and suggestions that
improved our paper significantly. We acknowledge the support of L.V.~T\'oth and
H.~Beuther in the ammonia measurements.
Based on observations with the Infrared Space Observatory, the Calar Alto 3.5\,m
telescope, the IRAM 30\,m, the Effelsberg 100-m telescope of the Max-Planck-Institut
f\"ur Radioastronomie, the James-Clerk-Maxwell Telescope and the Heinrich-Hertz-Telescope.
The ISOSS is supported by funds from the DLR, Bonn.
\end{acknowledgements}

\clearpage

\end{document}